\documentclass[12pt]{article}
\usepackage{graphics}
\usepackage{amssymb}
\usepackage{psfrag}

\newcommand{\rf}[1]{(\ref{#1})}
\newcommand{\beq}{\begin{equation}}
\newcommand{\eeq}{\end{equation}}
\newcommand{\bea}{\begin{eqnarray}}
\newcommand{\eea}{\end{eqnarray}}

\renewcommand{\a}{\alpha}

\newcommand{\del}{\delta}
\newcommand{\Del}{\Delta}

\newcommand{\ra}{\rangle}
\newcommand{\la}{\langle}

\newcommand{\Nof}{N_{14}}

\newcommand{\Ntt}{N_{23}}

\begin{document}

{\normalsize \hfill SPIN-2004/05}\\
\vspace{-1.5cm}
{\normalsize \hfill ITP-UU-04/11}\\
${}$\\ 

\begin{center}
\vspace{48pt}
{ \large \bf Emergence of a 4D World from  Causal Quantum Gravity}

\vspace{30pt}

{\sl J. Ambj\o rn}$\,^{a,c}$
{\sl J. Jurkiewicz}$\,^{b}$,
and {\sl R. Loll}$\,^{c}$

\vspace{24pt}
{\footnotesize

$^a$~The Niels Bohr Institute, Copenhagen University\\
Blegdamsvej 17, DK-2100 Copenhagen \O , Denmark.\\
{ email: ambjorn@nbi.dk}\\

\vspace{10pt}

$^b$~Institute of Physics, Jagellonian University,\\
Reymonta 4, PL 30-059 Krakow, Poland.\\
{ email: jurkiewicz@th.if.uj.edu.pl}\\

\vspace{10pt}

$^c$~Institute for Theoretical Physics, Utrecht University, \\
Leuvenlaan 4, NL-3584 CE Utrecht, The Netherlands.\\ 
{ email:  j.ambjorn@phys.uu.nl, r.loll@phys.uu.nl}\\

\vspace{10pt}

}
\vspace{48pt}

\end{center}


\begin{center}
{\bf Abstract}
\end{center}

Causal Dynamical Triangulations in four dimensions provide
a background-independent definition of the sum over geometries 
in nonperturbative quantum gravity, with a positive cosmological constant.
We present evidence 
that a macroscopic four-dimensional world emerges from this theory
dynamically.
 
\vspace{12pt}
\noindent


\newpage

\subsection*{Introduction}\label{intro}

String theory has highlighted the fact that the notion of ``dimension" 
in a quantum theory of all fundamental interactions has a very different 
status from that in any classical field theory or any perturbative quantum
field theory on a fixed background, where the dimension of spacetime
remains fixed throughout. This may be seen as a particular
case of the more general truth, not always appreciated, that in any 
nonperturbative theory of quantum gravity dimension will become a
dynamical quantity, along with other aspects of geometry. (By dimension
we mean an effective dimension observed at macroscopic scales.)

This Letter deals with an approach to quantum gravity
in which the dynamical property of spacetime dimension
is particularly transparent.\footnote{An interesting attempt to 
dynamically derive a four-dimensional spacetime in string theory 
has been made
in the nonperturbative M(atrix) formulation of type IIB strings
\cite{kawai1,kawai2,kawai3}, thus far with no definite conclusion on the
dimensionality \cite{nishimura1,nishimura2,nishimura3,nishimura4,
nishimura5,bielefeld,bielefeldagain}.}
The approach is that
of ``nonperturbative quantum gravity from Lorentzian dynamical
triangulations" or, for short, ``Causal Dynamical Triangulations". 
In it, the sum over all spacetime geometries takes the form of
a sum over four-geometries constructed from discrete building
blocks.\footnote{We would like 
to emphasize that the presence of discrete simplicial 
building blocks does not introduce any fundamental discreteness
of spacetime in the theory. The simplices are the analogues in
dynamical gravity of the fixed hypercubic
lattice cells used to define QCD nonperturbatively. The geodesic
length parameter $a$ characterizing their edge lengths
serves merely as a short-distance cut-off which is taken to zero in the 
continuum limit.} Each collection of building 
blocks, glued together according to a set of rules,
represents a geometry contributing to the path integral. 
To extract its  
physical geometric properties one has to perform the path integral,
renormalize, and evaluate the expectation values of appropriate 
observables, including the effective (or Hausdorff) dimension $d_h$.
 
Note that the dynamical nature of ``dimensionality" implies that
the Hausdorff dimension of the quantum geometry is {\it not} a priori
determined by the dimensionality at the cut-off scale $a$, which is
simply the fixed dimensionality $d$ of the building blocks of the
regularized version of the theory. An example in point are
the attempts to define theories of quantum geometry via ``{\it Euclidean} 
Dynamical Triangulations", much-studied during the
1980s and `90s. In these models, if the dimension $d$ is larger than 
2, and if all geometries contribute to the path integral with equal 
weight, a geometry with no linear extension and $d_h = \infty$ is created 
with probability one. If instead -- as is natural for a gravity-inspired
theory -- the 
Boltzmann weight of each geometry is taken to be the exponential of 
(minus) the Euclidean
Einstein-Hilbert action, one finds for small 
values of the bare gravitational coupling constant 
a first-order phase transition to a phase of the opposite 
extreme, namely, one in which the quantum geometry satisfies $d_h=2$. 
This is indicative of a different type of degeneracy, 
where typical (i.e. probability
one) configurations are so-called branched polymers or trees 
(see \cite{aj1,am,ckr,ds,aj2,bielefeld1,deb} for details of the phase
structure and geometric properties of the four-dimensional Euclidean
theory).
Unfortunately, this geometric degeneracy is also reflected in the
expectation values of other geometric quantities, and makes
Euclidean Dynamical Triangulations an 
unlikely candidate for a theory of four-dimensional quantum
gravity.

The requirement that a background-independent quantum gravity theory
possess the correct semi-classical limit, given by a macroscopically 
four-dimensional spacetime with microscopic quantum fluctuations,
is highly non-trivial.
The apparent failure of higher-dimensional Euclidean nonperturbative 
quantum gravity to do so (not limited to this particular approach \cite{livrev}),
together with a desire to formulate a quantum gravity with the correct
Lorentzian signature and causal properties \cite{teitelboim} were our 
main reasons for suggesting a radically different approach 
to the sampling of geometries in the path integral
\cite{al,ajl3d,ajl-prd,ajl4d}.
A further motivation was to construct a path integral more closely
related to canonical formulations of quantum gravity.
Starting from Lorentzian simplicial spacetimes with $d=4$ we insist 
that only causally well-behaved geometries appear in the 
(regularized) path integral, 
as described in detail in \cite{ajl4d}. 
A further crucial property of our explicit construction is that each 
configuration allows for a rotation to Euclidean signature, a
necessary prerequisite for discussing the convergence properties
of the sum over geometries, as well as for using Monte Carlo 
techniques.

In what follows we will report on the outcome of the first ever
Monte Carlo simulations of four-dimensional causal
dynamical triangulations. It differs radically from what was found
in previous simulations of four-dimensional {\it Euclidean}
dynamical triangulations. We will present strong evidence that
the Lorentzian framework produces a quantum
geometry which is both extended and effectively four-dimensional.
This is to our knowledge the first example of a theory of
quantum gravity that generates a quantum spacetime with
such properties {\it dynamically}.

\subsection*{The emergent macroscopic 4d geometries}

All causal simplicial spacetimes contributing to the path integral
are foliated by an integer-valued ``proper time" $t$, and each geometry
can be obtained by gluing together four-simplices 
in a way that respects this foliation. Each four-simplex has timelike
links of length-squared $a_t^2$ and spacelike links
of length-squared $a_s^2$, with all of the latter being located in spatial
slices of constant integer-$t$. These slices consist of purely
spacelike tetrahedra, forming a three-dimensional piecewise flat
manifold, whose topology we choose to be that of a three-sphere
$S^3$.

Each spatial tetrahedron at time $t$ is 
shared by two four-simplices (said to be of type (1,4) and (4,1)) 
whose fifth vertex lies in the neighbouring
slice of time $t-1$ and $t+1$ respectively. In addition
we need four-simplices of type (2,3) and (3,2) which share one link and one 
triangle with two adjacent spatial slices
\cite{ajl4d}. 
Let us assume the link lengths
are related by
\beq\label{1.1}
a_t^2 = -\a a_s^2.
\eeq
All choices $\a > 0$ correspond to Lorentzian and all choices $\a < -7/12$
to Euclidean signature, and a Euclideanization of geometry amounts
to a suitable analytic continuation in $\alpha$.
Setting $\a = -1$ leads to a particularly 
simple expression for the (Euclidean) Einstein-Hilbert action of a given 
triangulation $T$ (since all four-simplices
are identical geometrically), namely, 
\beq\label{2}
S(T) = -k_0 N_0(T) + k_4 N_4(T),
\eeq
with $N_i(T)$ denoting the number of $i$-dimensional simplices in
$T$. In \rf{2}, $k_0$ is proportional 
to the inverse (bare) gravitational coupling 
constant, $k_0 \sim 1/G$, while $k_4$ is a 
linear combination of the cosmological and inverse gravitational 
coupling constants. The action \rf{2} has been calculated 
from Regge's prescription for piecewise linear geometries.
If we take $\a \neq -1$ the Euclidean 
four-simplices of type (1,4) and type (2,3)
will be different and thus appear with different weights in the 
Einstein-Hilbert action \cite{ajl4d}. For our present purposes 
it is convenient to use the equivalent parameterization
\beq\label{3}
S(T) = -k_0 N_0(T) + k_4 N_4(T) + \Del (2\Nof(T)+\Ntt(T)),
\eeq
where $\Nof(T)$ and $\Ntt(T)$ denote the combined numbers
in $T$ of four-simplices
of types $(1,4)$ and $(4,1)$, and of types
$(2,3)$ and $(3,2)$, respectively.
The explicit map between the parameter $\Del$ in eq.\rf{3} and $\a$ 
can be readily worked out. For 
the simulations reported here we have used $\Del$'s in the range 0.4--0.6.

\begin{figure}[h]
\centerline{\scalebox{1.1}{\rotatebox{0}{\includegraphics{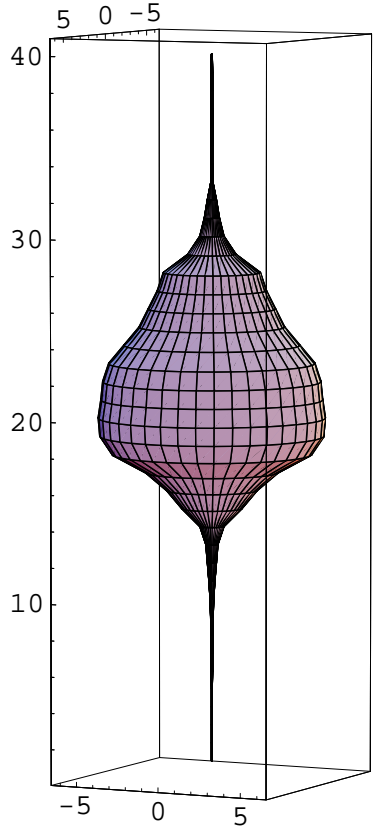}}}}
\caption[phased]{{\small 
Snapshot of a ``typical universe'' of volume 91.1k from
the Monte Carlo simulations. The total time extent (vertical direction)
is $t=40$ as indicated. The circumference at integer-$t$
is the spatial three-volume $V_3(t) \times 0.02$ (in 
units where $a_s=1$).
The surface represents an interpolation between 
adjacent ``spatial volumes''. No attempt has been made to capture the 
actual 4d connectivity between neighbouring spatial slices.}}
\label{fig1}
\end{figure}

The quantum-gravitational proper-time propagator is defined by
\beq\label{4}
G_{k_0,k_4,\Del}( T^{(3)}(0),T^{(3)}(t)) = 
\sum_{T_t} \frac{1}{C_{T_t}} \; e^{-S(T_t)}. 
\eeq
where the summation is over all four-dimensional 
triangulations $T_t$ of topology $S^3\times [0,1]$ and $t$
proper-time steps, whose spatial 
boundary geometries at proper times $0$ and $t$ are
$T^{(3)}(0)$ and $T^{(3)}(t)$.
The order of the automorphism group of the 
triangulation $T_t$ is denoted 
by $C_{T_t}$. The propagator can be 
related to the quantum Hamiltonian conjugate to $t$, and
in turn to the transfer matrix of the (Euclidean) statistical theory
\cite{ajl4d}. 

While it may be difficult to find an explicit analytic 
expression for the full propagator \rf{4} of the four-dimensional theory,
Monte Carlo simulations are readily available, employing standard 
techniques
from Euclidean dynamically triangulated quantum gravity
\cite{MC}.  Ideally one would like to keep the  
{\it renormalized} 
cosmological constant $\Lambda$ fixed in the 
simulation.\footnote{For the relation between the bare 
(dimensionless) cosmological constant $k_4$ and $\Lambda$ see \cite{aj1}.} 
The presence of the cosmological term $\Lambda \int \sqrt{g}$ in the action then 
implies that the four-volume $V_4$ fluctuates around 
$\la V_4 \ra \sim \Lambda^{-1}$. However, 
for simulation-technical reasons one fixes instead
the average number $\la N_4 \ra$ of four-simplices,
or $\la V_4\ra$,  
from the outset\footnote{For fixed $\a$ (or $\Delta$) one has 
$\la \Nof \ra \propto \la \Ntt \ra \propto \la N_4 \ra$. According to \cite{ajl4d} 
we have $V_4 = a_s^4(\Nof\sqrt{8\a+3} + \Ntt\sqrt{12\a +7}).$
We set $a_s=1$.}, effectively 
working with a cosmological constant $\Lambda \sim \la V_4 \ra^{-1}$.

\begin{figure}[ht]
\vspace{-3cm}
\psfrag{VV}{\Large $VV(\delta)$}
\psfrag{delta}{\Large $\delta$}
\centerline{\scalebox{0.6}{\rotatebox{0}{\includegraphics{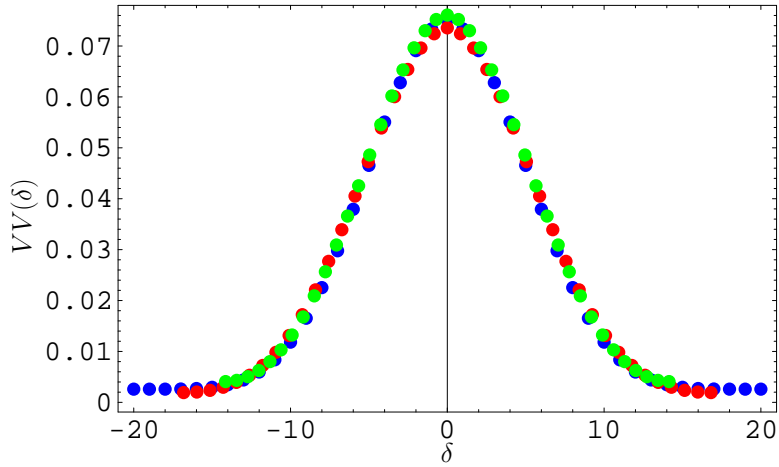}}}}
\vspace{-4.5cm}
\caption[phased]{{\small Measurement of spatial volume-volume correlators 
for spacetimes with  $N_4 \approx$ 45.5, 91.1 and 
184.4k, 
and after performing the rescaling \rf{5}.}}
\label{fig3}
\end{figure}

Fig.\ref{fig1} shows a snapshot of a configuration with
$N_4 \approx$ 91100 four-simplices
as it appears in the Monte Carlo simulations. 
Important information is contained in how the volume $V_3$ of
spatial slices and the time extent $\tau$ of the observed 
universe\footnote{The time extent $\tau\leq t$ measures the time
during which the universe has spatial slices of macroscopic 
size $V_3\gg 1$.} 
behave in relation to the total spacetime volume $V_4$.
To good 
approximation we find that the spatially extended parts of 
the spacetimes for various four-volumes $V_4$ can be
mapped onto each other by rescaling the spatial volumes
and the proper times according to
\beq\label{5}  
V_3 \to V_3/V_4^{3/4},~~~~\tau \to \tau/V_4^{1/4}.
\eeq
This is illustrated in Fig.\ref{fig3}, displaying the so-called volume-volume
correlator 
\beq\label{v-v}
\la V_3(0)V_3(\del)\ra = \frac{1}{t^2} \sum_{j=1}^{t} V_3(j)V_3(j+\del)
\eeq
measured for three different spacetime volumes $V_4$, using the rescaling 
\rf{5}, and where we have periodically identified $T_3(t)=T_3(0)$. The
detailed statistical analysis leading to (\ref{5}), (\ref{v-v}) and the measurement
of the spatial Hausdorff dimension below will be published elsewhere 
\cite{ajl-appear}
(see also \cite{ajl3d} for an analogous treatment in three dimensions).

\begin{figure}[ht]
\vspace{-3cm}
\psfrag{log2}{\bf{\Large $\log(\langle r\rangle + 1.75)$}}
\psfrag{log1}{\Large\bf $\log(\langle V\rangle)$}
\centerline{\scalebox{0.6}{\rotatebox{0}{\includegraphics{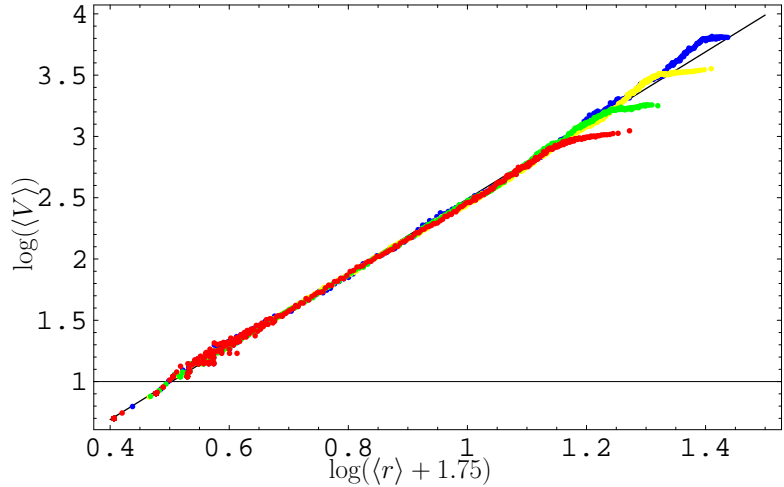}}}}
\vspace{-4.5cm}
\caption[phased]{{\small The log-log plot 
of the average geodesic distance $\la r\ra_{V_3}$ versus $V_3$.
It combines data from universes where  $N_4 \approx$ 
45.5, 91.1, 184.4 and 371.2k.}}
\label{fig2}
\end{figure}

Relation \rf{5} strongly suggests that what we ``observe'' 
in the simulations are genuine four-dimensional universes whose spatial 
hypersurfaces are three-dimensional. To test 
this we have recorded the average geodesic distance $\la r\ra_{V_3}$ between 
points in the spatial three-volumes $V_3$. If the spatial geometry 
is indeed three-dimensional one expects a leading behaviour
\beq\label{6}
\la r\ra_{V_3} \sim V_3^{1/3}.
\eeq 
Fig.\ref{fig2} shows the results of measuring
$\la r \ra_{V_3}$ in the macroscopic spatial slices of
our quantum universes.\footnote{The appearance of a small additive 
constant $\la r\ra \to \la r \ra +c$, $c=1.75$, is a finite-size scaling 
effect familiar from earlier studies of Euclidean quantum gravity
\cite{ajw,aj2} which improves the range of applicability of \rf{6} for
small $V_3$.} It combines the data 
for universes 
with total  $N_4 \approx$  45.5, 91.1, 184.4 and 371.2k.
The straight line in Fig.\ref{fig2} has merely been drawn to guide the eye;
it corresponds to the three-dimensional 
behaviour \rf{6}. From these combined
measurements, the evidence for {\it three}-dimensional slices is rather
compelling.
A best fit for the spatial Hausdorff dimension $d_h$ in the relation 
$\la r\ra_{V_3} \sim V_3^{1/{d_h}}$
yields $d_h = 3.10 \pm 0.15$. As should be
clear from the Introduction, reproducing this ``classical'' dimensionality
dynamically from a fully nonperturbative formulation of quantum gravity
constitutes a highly nontrivial result. However, it should be kept in mind that a
property like $\la r \ra_{V_3} \sim V^{1/3}_3$ provides only a crude 
characterization of the spatial geometry, and by no means implies
that the space ``observed" in the computer simulations is a nice smooth 
3d space at short distances. --
Further details about the simulations, measurements,
fits, and the complete phase diagram of the model will be 
published shortly \cite{ajl-appear}. 

\subsection*{Discussion}

Causal dynamical triangulations are a framework for defining
quantum gravity nonperturbatively as the continuum
limit of a well-defined regularized sum over
geometries. Interestingly, and in complete agreement with current
observational data is the fact that the physical cosmological constant
$\Lambda$ in dynamical triangulations is necessarily positive.\footnote{Note 
that without further observational input the value of $\Lambda$ remains
a free parameter of the quantum theory.}  

Recall that the effective cosmological constant in 
our simulation is $1/\la V_4 \ra$. 
Since, according to \rf{5}, both the spatial three-volumes and the time 
extension of 
our quantum universe relate to $\Lambda$ canonically 
(that is, in a way expected from their classical dimensionality), 
it is appropriate to call this universe macroscopic.   

This leads us to conclude that we have observed the
emergence of a four-dimensional macroscopic world with 
three-dimensional spatial geometries. Judging from the computer
simulations, this dynamically generated quantum geometry 
acts as a background geometry 
around which small quantum fluctuations take place.
Further numerical studies will be needed to make this into
a quantitative statement. The situation is really rather
remarkable: we started out from an explicitly background-independent
formalism and have rederived a particular stable background
geometry. Obviously, questions about the nonrenormalizability of
perturbative gravity
will have to be readdressed in this new context (maybe along
the lines outlined in \cite{ajl5}), with the benefit
of having an explicit microscopic model of quantum spacetime
and its quantum fluctuations. 
We hope future simulations and analytic studies of the model will 
teach us how to reconcile these aspects.
The final picture may be that of
an asymptotically safe theory in the sense of Weinberg \cite{as},
or turn out to require a completely new theoretical framework.

\subsection*{Acknowledgment}
All authors acknowledge support by the
EU network on ``Discrete Random Geometry'', grant HPRN-CT-1999-00161. 
In addition, J.A.\ and J.J.\ were supported by ``MaPhySto'', 
the Center of Mathematical Physics 
and Stochastics, financed by the 
National Danish Research Foundation.
J.J.\ acknowledges support by the Polish Committee for Scientific Research 
(KBN) grant 2P03B 09622 (2002-2004).

\end{document}